# Extremely broadband ultralight thermally emissive metasurfaces

Ali Naqvi[1], Samuel P. Loke[1], Michael D. Kelzenberg[1], Dennis M. Callahan[1], Emily C. Warmann[1], Pilar Espinet-González[1], Nina Vaidya[1], Tatiana A. Roy[1], Jing-Shun Huang[1], Tatiana G. Vinogradova[2], Alexander J. Messer[2], Harry A. Atwater[1]

[1]California Institute of Technology, Pasadena, CA 91125

[2]Northrop Grumman Aerospace Systems, 1111 W 3rd St, Azusa, CA 91702

Abstract

We report the design, fabrication and characterization of ultralight highly emissive metaphotonic structures with record-low mass/area that emit thermal radiation efficiently over a broad spectral (2 to 35 microns) and angular (0–60º) range. The structures comprise one to three pairs of alternating nanometer-scale metallic and dielectric layers, and have measured effective 300 K hemispherical emissivities of 0.7 to 0.9. To our knowledge, these structures, which are all subwavelength in thickness are the lightest reported metasurfaces with comparable infrared emissivity. The superior optical properties, together with their mechanical flexibility, low outgassing, and low areal mass, suggest that these metasurfaces are candidates for thermal management in applications demanding of ultralight flexible structures, including aerospace applications, ultralight photovoltaics, lightweight flexible electronics, and textiles for thermal insulation.

1. Introduction

Understanding the limits to far field [1,2] radiative energy transfer is fundamental to many areas of science. This subject has enjoyed expanded interest and effort in recent years with the advent of tailored electromagnetic materials such as metamaterials and metasurfaces [3-5]. Fundamental interest in the far field thermal radiation is multifold. On one hand, it is the primary way of cooling objects in space [6], where convection is lacking. On the other hand, it provides a means to manipulate optical forces in space via radiation pressure management [7]. Besides, appropriate control of far field thermal radiation can increase the efficiency of solar cells [8], thus can have a global impact.

From a practical viewpoint, tailoring the emission of thermal radiation has the potential to benefit a wide range of applications including radiative cooling for terrestrial use [9], thermophotovoltaic energy harvesting [10], optoelectronics and plasmonics of two-dimensional materials [11], heat-assisted magnetic recording [12], cooling of nanoscale electronic devices [13], developing reconfigurable optical platforms and devices [14], sensing [15], and thermal management of aircraft and spacecraft [6]. Specifically, realizing structures that efficiently emit infrared radiation in the 8 to 14 micron range can enable passive thermal management of devices operating near room temperature, as this spectral range corresponds to the peak of the blackbody spectrum at a temperature of 300°K-400°K. Appropriate radiative cooling at this temperature range benefits applications including the efficient performance of photovoltaic cells, metabolism of living organisms and thermal signature control.

A key factor motivating our work is the design and realization of thermally emissive structures that are as lightweight as possible, a feature that is important for space-based technologies where design of active structures with lowest possible mass per unit area is a critical metric. For a homogeneous dielectric medium, reducing the thickness also inevitably reduces the optical absorptivity and emissivity, particularly as the thickness is reduced to less than several wavelengths. Thus the challenge for ultralight metasurfaces is to achieve high emissivity at subwavelength structural thicknesses. Metasurface designs have promise to minimize the emitter mass, as they enable one to manipulate electromagnetic wave amplitude, phase and polarization over subwavelength dimensions [16,17]. Optical frequency metasurfaces have recently been a subject of widespread scientific interest and have found application in diverse areas of research including implementation of flat optics platforms [18], beam steering for active imaging applications [19] and optical computation [20].

Conventional thermally emissive structures include a broad range of particulate and bulk materials and composites. Black or white paints and different types of materials such as anodized metals, carbon fiber and carbon nanotubes [21] are widely used in current thermal radiation applications. Despite their ease of use, application of paints as emissive materials is limited in a practical sense as their thickness cannot be typically made thinner than about several tens of microns [22], i.e., a thickness of several infrared wavelengths. Fundamentally, the absorbance and emittance decreases correspondingly as the layer is made thinner than a few wavelengths. Metals (e.g., Al, Cr, Ag and Au) are characterized by a large imaginary permittivity component at infrared wavelengths and have potential for high infrared emissivity. However, the drawback of homogeneous metallic structures is that they have relatively high areal mass densities and are also very reflective. Polymeric organic materials such as polyimides are among other alternatives that are very lightweight and show strong vibronic resonances at wavelengths less than 10 microns, however, they are weakly absorptive and emissive at long wavelengths [23], and typically require thicknesses of tens of microns to achieve high effective emissivity in the 300-400° K range. To manipulate thermal radiation in structures made of thin polymers, nanophotonic design can be used to alter effective optical properties over the thermal infrared wavelength range.

Here, we report the design, fabrication and characterization of polymeric thermally emissive metasurfaces based on Salisbury screen [24,25] and Jaumann absorber [26] concepts. The Salisbury screen is a widely-employed electromagnetic wave absorber that consists of a quarter-wavelength dielectric layer placed between a metallic back reflector and a thin conductive sheet. The high absorption by the Salisbury screen can be explained by destructive interference of the incident and the reflected waves [27,28] and therefore, depends on the incident wavelength and angle. The absorption bandwidth of Salisbury screens can be increased by adding additional dielectric-metallic bilayers, yielding multilayered structures that have been termed Jaumann absorbers [29,30]. Thus the Salisbury screen is the simplest form of a Jaumann absorber, and one can view it as a Jaumann absorber with a single layer pair on the back reflector. We realize our metasurfaces using polyimides as the dielectric, and thin metallic sheets that are much thinner than the optical skin depth for thermal infrared radiation.

Finally, in our design and measurement procedures, we use the reciprocity relations between the emission and absorption inherent to Kirchhoff's thermal radiation law, which dictates that the absorptance and emittance of a surface are equal at a given wavelength, angle and polarization [31]. Thus, emissivity data presented herein is based on measurements or calculations of absorption.

## 2. Design of the emissive metasurfaces

In our designs, we utilized a back reflector, dielectric spacer and top metallic sheet comprised of Cr, CP1 polyimide and Cr respectively (see experimental for details). Optimization of the top metallic sheet and dielectric layer thicknesses by rigorous calculations leads to emissivity values depicted in Figure 1 (a). Two high-emissivity regions exist which correspond to small and large dielectric thicknesses. High emissivity at large dielectric thickness is expected because the dielectric can absorb and emit thermal radiation more efficiently over large thicknesses according to the Beer-Lambert law. In contrast, the high emissivity region at small dielectric thicknesses is due to interference effects that are the basis of a Salisbury screen. This phenomenon is more clearly demonstrated in Figure 1 (b), which shows the spectral emissivity as the Cr layer thickness changes and for a polyimide layer thickness fixed at its optimal thickness. *Spectral emissivity* is defined as the ratio of the power emitted from the surface at a particular wavelength to the power emitted from the blackbody at the same wavelength. The high-emissivity region at a wavelength near 10 microns and Cr thickness of 2 nm is the signature of the mentioned interference. The sharp peaks in emissivity in the range from 5 to 10 microns are related to vibronic resonances related to molecular vibrational modes of the polyimide material. To achieve a design that minimizes areal mass density, we find optimal thicknesses of 2.1 microns and 2 nm for the polyimide and the Cr layer respectively, which corresponds to an emissivity of 0.65 and an areal mass of 3.3 g/m$^2$.

An interesting question is whether we can replace Cr by other metals. Changing the back reflector material to Al, Ag, or Au has negligible effect on the results. However, the optimization of the front metal layer thickness depends strongly on the optical properties of the metal used. Figure 1(c) shows the emissivity optimization assuming the top metallic layer is made of Al. Notably, high emissivity values can still be obtained, but only if the Al sheet is dramatically (~10x) thinner than the optimal thickness range for Cr. Furthermore, the emissivity values are very sensitive to the Al layer thickness and drop dramatically with slight changes in Al thickness, as the Al quickly becomes opaque and reflective with increasing thickness. The reason can be found by comparing the optical permittivities of the two metals. Figure 1(d) illustrates the real and imaginary parts of the relative permittivity of Cr [32] and Al [33]. These values are about an order of magnitude smaller for Cr than for Al; thus, a much thinner layer of Al is required to induce a given amount of phase or amplitude change in the wave front.

Interestingly, for both Cr and Al structures, the optimal emissivity obtained is about 0.65. However, we note that the optimal thickness predicted for Al is on the order of the atomic spacing, and thus our calculations based on bulk optical properties may be inadequate. Furthermore, fabrication of such thin layers would likely require advanced methods, especially considering the tendency of bare Al to oxidize. We therefore conclude that Cr is suited for experimental realization of these structures, due to its low relative permittivity among metals, and because it can be readily deposited at optimal thickness using physical vapor deposition (e.g., electron-beam evaporation as employed in this work). Although exposed Cr surfaces also oxidize, like Al, the process is self-limiting and extends only a few monolayers from the surface, which is much less than the desired Cr thickness range. To further limit the effects of oxidation or other chemical reactions with the Cr, in our experimental work, we follow each Cr deposition by depositing a ≥10 nm layer of $SiO_2$, without breaking vacuum. This layer only slightly affects the optical behavior of the metasurface, but is included in the remainder of our calculations and measurements below.

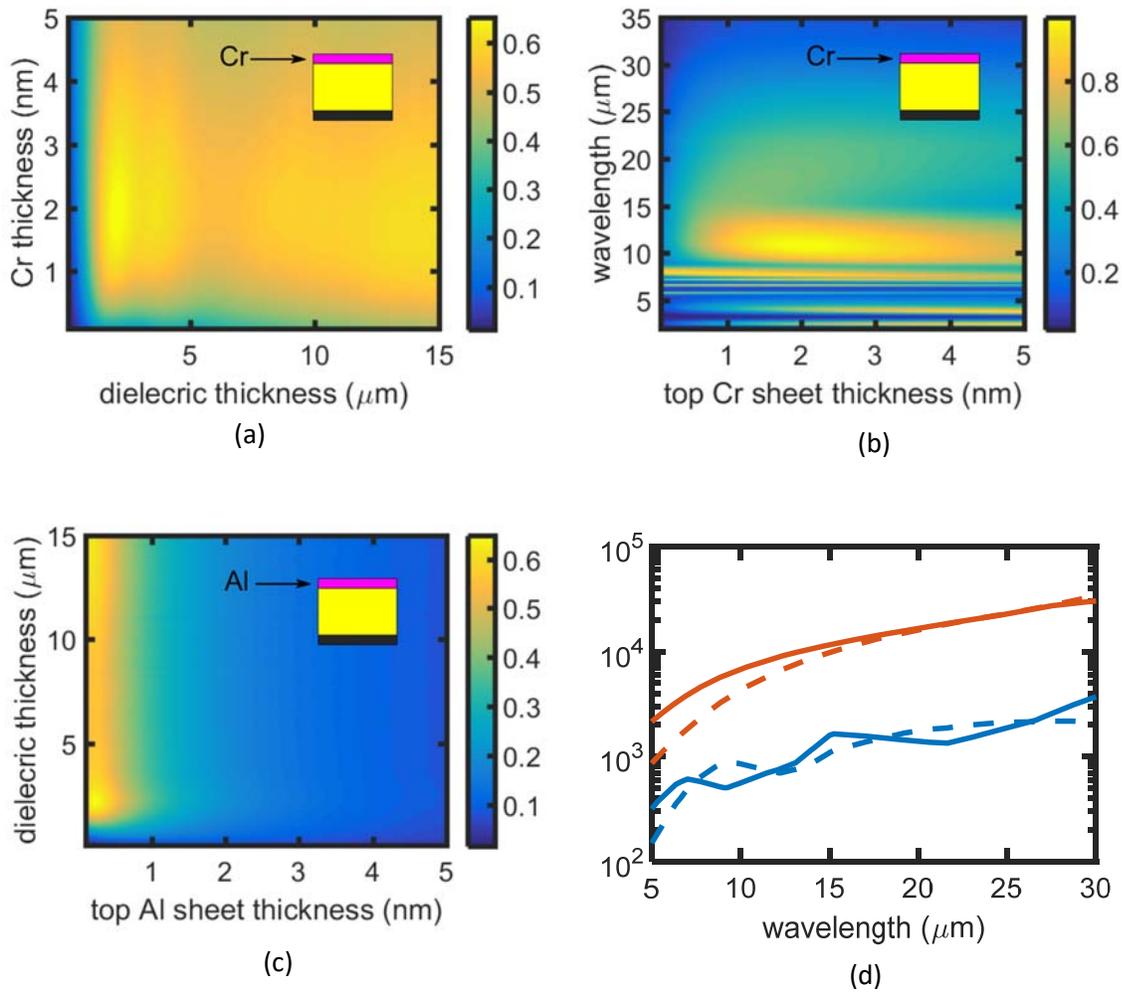

Figure 1- (a)The emissivity (300°K) of the Salisbury screen versus the Cr and the polyimide layer thickness. (b): Spectral emissivity of the Salisbury screen as the Cr layer thickness changes, for dielectric thickness of 2.1μm. Note that the scale is different from (a). (c): Same as (a) except that the top Cr sheet is replaced by an Al sheet. (d): Magnitude of the real and imaginary part of the relative permittivity of Cr [32](blue) and Al [33] (red). The real and imaginary part of the permittivity are indicated by the solid and dashed lines respectively. In (a) to (c) the top pink layer

*is the metallic sheet made of either Al or Cr, the yellow spacer is the CP1 layer and the bottom black layer is the back reflector.*

We next consider multilayer structures (Jaumann absorbers) in pursuit of even higher emissivity. Using the above approach, we calculated optimal layer thicknesses for structures having one, two, and three Salisbury screen layer pairs. It is instructive to analyze the fraction of thermal emission arising from the different layers in these structures. Figure 2 shows the layer-by-layer contribution to the spectral emissivity at normal incidence for the three optimized emissive surfaces. For clearer illustration, the emission spectra are plotted in a cumulative fashion. The top-most curve shows the total spectral emissivity at normal incidence and the shaded areas between successive curves show the proportional contribution of each corresponding layer. The additional $SiO_2$ layers are present to avoid interfacial reactions in the layered structure. In all cases, the very thin Cr layers dominate the emission, particularly at long wavelengths. This is in agreement with the previous calculations that show strong localization of the electromagnetic field in the metallic sheet [25]. In all cases, the back reflector emission is negligible.

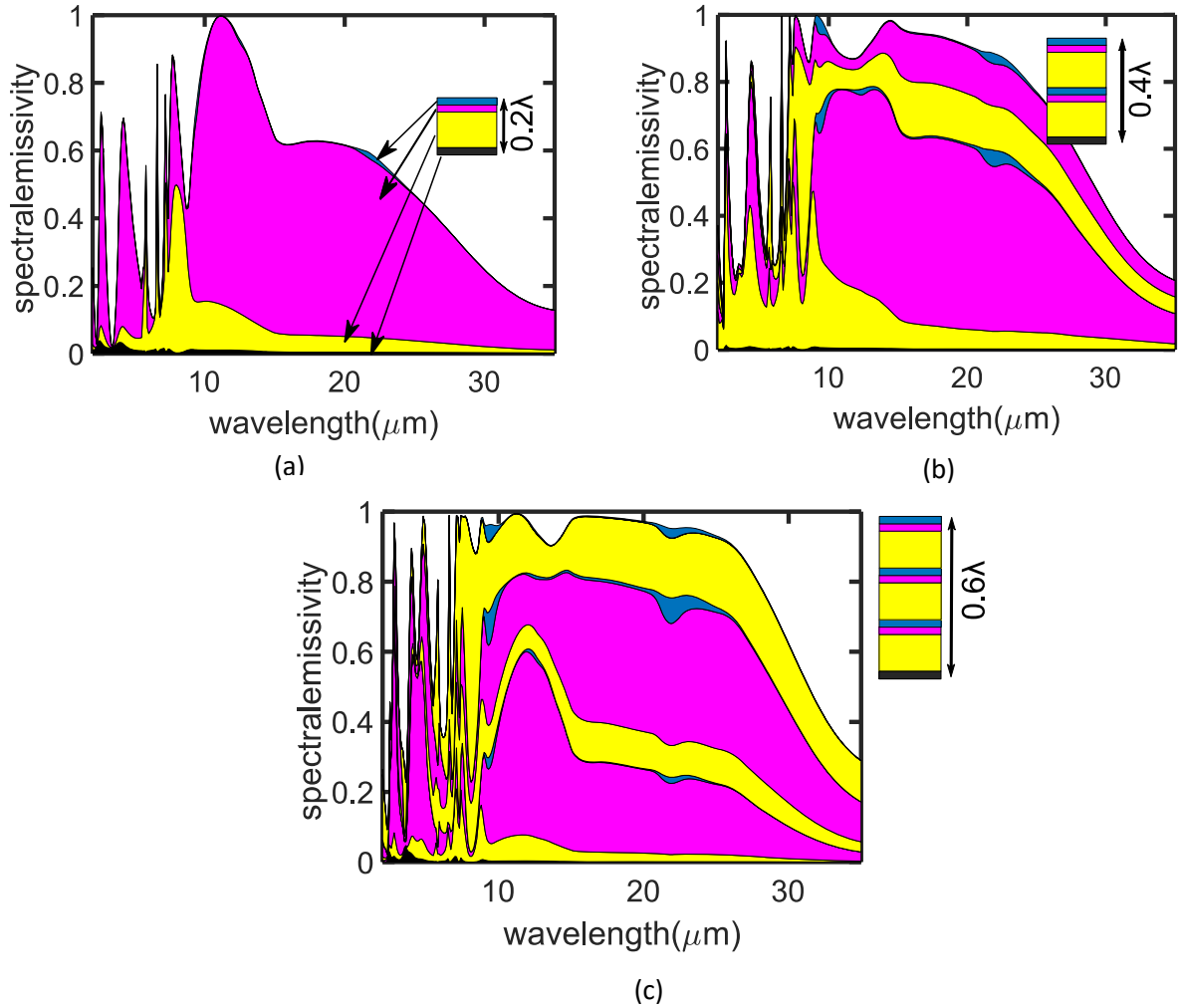

*Figure 2- The spectral emissivity that occurs in different layers in the (a): Salisbury screen with the thickness of ~2μm, (b): two-layer Jaumann absorber with the total thickness of ~4μm, and (c): three-layer Jaumann absorber with the total thickness of ~6μm. In all cases the emissivity of the layers is plotted cumulatively, in the same order as the layers are shown. The $SiO_2$, Cr and polyimide layers are shown in blue, pink and yellow, respectively. The back reflector is indicated in black. The values are obtained from rigorous calculations of absorption.*

Until now, we considered only emission in the direction normal to the metasurface. To have an accurate estimate of the total power that is emitted by the metasurface at each temperature, the angular dependence of thermal radiation should also be taken into account. Figure 3 (a) shows the *directional emissivity*—which is the weighted average of spectral emissivity by the blackbody spectral radiance weighting—versus the emission angle for all three designed structures, in both TE and TM polarization. The TM-polarized emissivity is in all cases larger than the TE-polarized emissivity due to the Brewster effect, which occurs close to 70 degrees. The unpolarized directional emissivity is almost constant over a large angular range, from normal to 70 degrees, as the drop in the TE-polarized emissivity is compensated by the TM-polarized emissivity. The total power radiated from any of the planar structures can be obtained by applying the Lambert cosine law. Therefore, the *hemispherical emissivity*, which is the total power radiated from the structure at all angles, to the total power radiated from a blackbody can be obtained by

$$\overline{\epsilon_\theta} = \frac{P_{rad}}{P_{rad,BB}} = \frac{\int d\Omega \cos\theta \int d\lambda I_{BB}(\lambda,T)\epsilon(\lambda,\Omega)}{\int d\Omega \cos\theta \int d\lambda I_{BB}(\lambda,T)}$$

where $\lambda$ is wavelength, $T$ is temperature, $\Omega$ is solid angle and $I_{BB}$ is the blackbody spectral radiance. Figure 3 (b) shows the emissivity of the structures versus their areal mass for both normal and hemispherical emission. Increasing the number of layers yields a higher emissivity and a correspondingly larger areal mass. The emissivity of the three-layer structure reaches 90% (normal) and 84% (hemispherical). Even for the simplest structure—a Salisbury screen—, a high emissivity value of 65 % is calculated for emission in the normal direction. The areal mass of each of these emissive surfaces is less than 10 g/m². Specifically, the Salisbury screen weighs only 3.3 g/m², which is to our knowledge, the lightest structure with this level of emissivity. Here, the mass of the back reflector is excluded from the calculation of areal mass, because we consider the metal as an existing surface whose emissivity we seek to increase by addition of the metasurface layers. Should we instead desire to fabricate a free-standing membrane, or to modify the emissivity of a transparent surface, we could add a 40 nm Al layer beneath the dielectric, increasing the areal mass by 0.1 g/m².

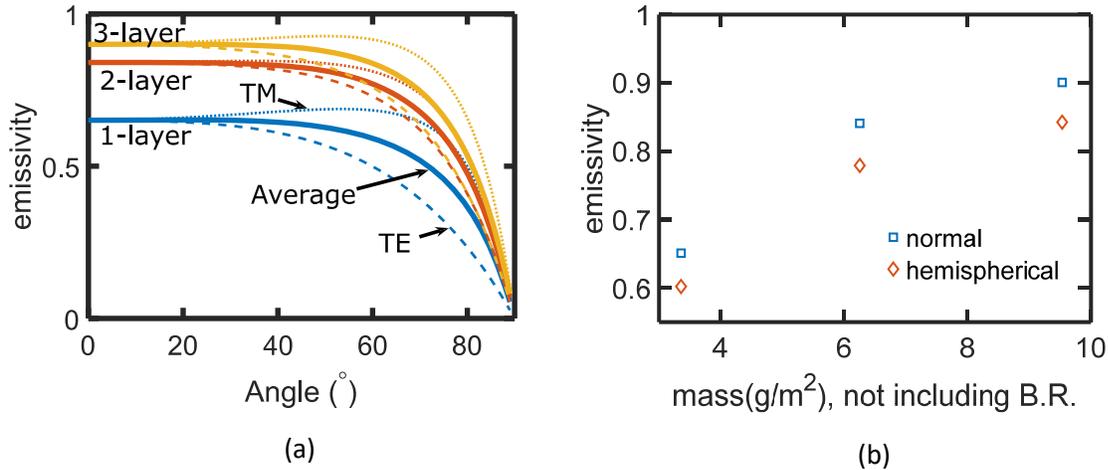

*Figure 3- (a): Directional emissivity for both TE- and TM-polarized emission for the three studied structures. TE, TM and unpolarized values are shown by the dashed, dotted and solid curves, respectively. Blue, red and yellow refer to the 1-layer, 2-layer and 3-layer structures. (b): Emissivity vs. areal mass for the three studied structures. The squares and the diamonds show the emissivity at normal angle (squares) and hemispherical (diamonds). The back reflector mass is excluded.*

A reasonable question is why such wideband spectral emissivity is predicted for even the single-layer structure, despite the general narrowband nature of the Salisbury screens. As expected, in Figure 2, the emissivity for the Salisbury screen is large at a wavelength near 10 microns because this wavelength is close to the blackbody spectrum peak at around 300 °K. However, the associated resonance falls off rapidly as the wavelength increases. The broad peak at 20 microns should exist because the permittivity of Cr becomes very large at longer wavelengths (c.f. Figure 1 (d)). See Figure S1 for the details of the angular dependence of the spectral emissivity in different polarizations.

Figure 4 (a) shows a SEM image of the cross section of a Salisbury screen that was fabricated. Due to the multiscale dimensions of the structure, the Cr layer and the $SiO_2$ layer are observed as one very thin sheet on top. Figure 4 (b) compares the infrared spectral emissivity of the Salisbury screen that is obtained from two different measurements with the spectroscopic ellipsometer and Fourier transform infrared (FTIR) microscope, which are in excellent agreement. (See Figure S2 for similar plots of two- and three-layer structures, and angle-resolved spectra for all three samples. The angle-resolved data presented here represent the average of TE and TM polarization measurements.) Weighting by the 300 K blackbody spectrum and integrating over wavelength gives the directional emissivity of each sample versus angle, which is depicted in Figure 4 (c). The thermal emissivity of all three structures is more than 0.7 for angles up to almost 60 degrees. Expectedly, as the number of layers increases, the emissivity is enhanced, especially at long wavelengths and at shallower angles.

Interestingly, the reflectance measurements for all three samples indicates slightly higher emissivity than predicted by the calculations and optimizations above. Different processes are responsible for the large emissivity of these metasurfaces. The thin Cr layer is not a uniform layer and consists of small nanoparticles with various sizes. For metal nanoparticles with nanometer-sized diameters, the energy levels in the conduction band become discretized due to quantum size effects [34]. Besides, the polyimide layer has a big molecular composition and supports a considerable number of vibronic resonances. As electrons in the thin Cr layer get hot, they give some part of their energy through thermalization to the Cr lattice and subsequently to the surrounding media. The energy of the hot electrons at the metal-polyimide interface can get coupled to the vibronic and phononic resonances of the polyimide layer. We believe this coupling is responsible for a major part of the absorption/emission enhancement in our metasurfaces. Although transfer of energy of hot electrons to the $SiO_2$ lattice may have a marginal role in emission, we speculate that the top $SiO_2$ layer cannot be responsible for the very broadband emission in the IR range because it has phonon resonances with much narrower widths.

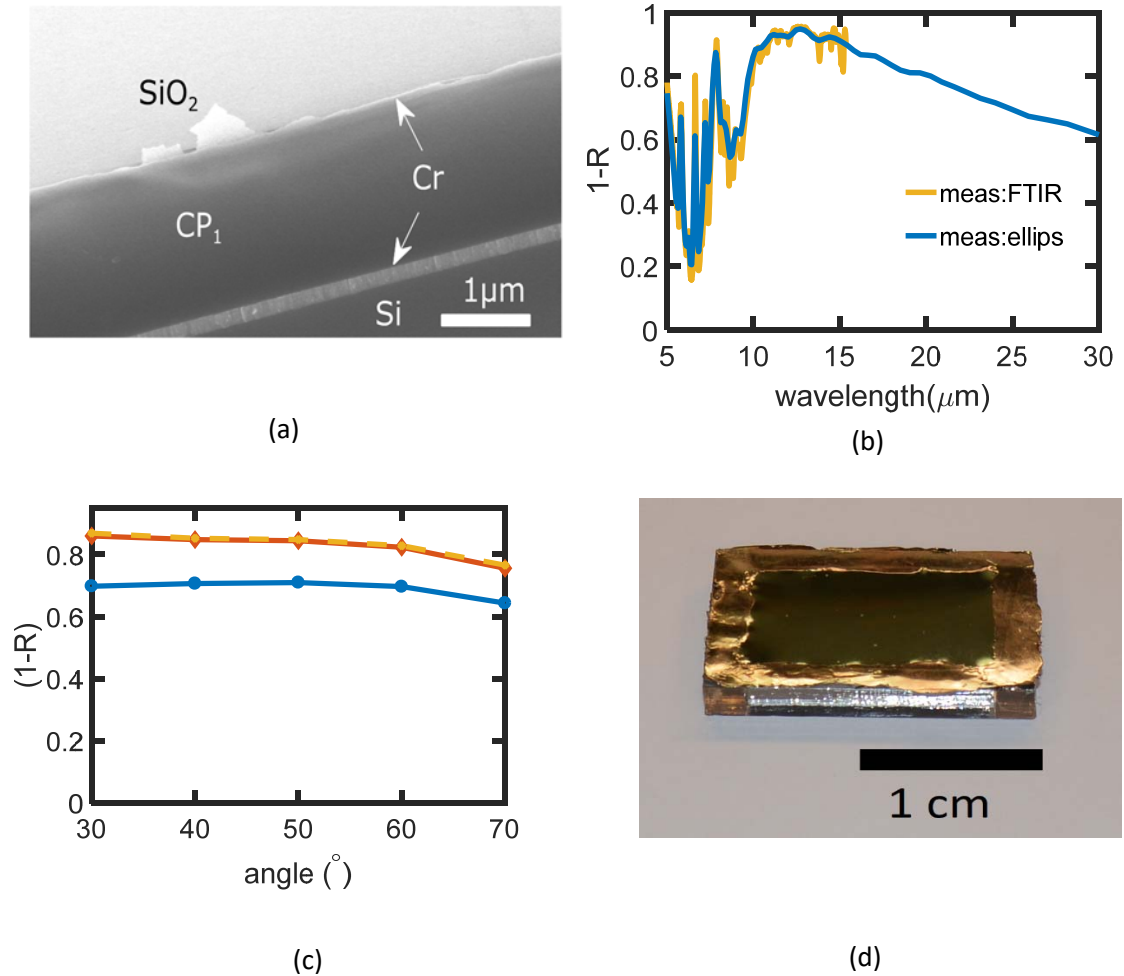

*Figure 4- (a): The SEM micrograph of the fabricated Salisbury screen. (b): Infrared spectral absorption of the Salisbury screen as obtained by FTIR (yellow) and ellipsometer (30° incidence, blue). The measurements were done with a Nicolet iS50 FTIR coupled to a Continuum microscope with a 100 µm spot size (c): The angle-dependent emissivity for the three fabricated metasurfaces. (d) The fabricated free-standing Salisbury screen installed on a frame. The flat central part is the Salisbury screen with a total thickness of around 2.1 µm. The surrounding parts meet the underneath frame, hence have a different shape.*

We also fabricated a free-standing Salisbury screen membrane, shown in Figure 4 (d). The 300 K emissivity of this sample was 0.60 as inferred from spectroscopic reflectance measurements at 30° incidence angle (see Figure S3). This value is slightly less than that obtained for the Salisbury screen that was fabricated on a rigid substrate (above). The primary reason for this difference is that the free-standing membrane has a dielectric thickness of 1.8 µm, which is slightly less than optimal (see Figure 1). The rigid structure fabricated above used the optimal dielectric thickness of 2.1 µm.

The high emissivity of the free-standing membrane is nonetheless remarkable considering that its total thickness is less than 2 µm, corresponding to a calculated areal mass of 3 g/m$^2$. For comparison, the aluminized polyimide (Kapton) sheeting typically used in spacecraft multilayer insulation must be 25 microns thick (36 g/m$^2$) in order to achieve a similar value of emissivity (0.62) [35]. Although working with such thin membranes may be challenging in terms of fabrication, handling, and impact damage, the ability to achieve moderate to high emissivity with 3–10 g/m$^2$ areal mass may be beneficial to numerous aerospace applications. And, although achieving high emissivity with these structures requires precise control of dielectric layer thickness as well as extremely thin layers of Cr metallization, this is well within the capabilities of established roll-to-roll fabrication processes. Thus, ultralight high-emissivity films can be prepared and subsequently integrated or laminated onto other surfaces, enabling the thermal emittance of a structure to be designed and manufactured separately from its other functions.

3. Experimental

**Fabrication.** We fabricate the Salisbury screen by evaporating a 100 nm thick Cr back reflector layer on a Si substrate, followed by spin coating the Nexolve CP1 [36] polyimide layer, then electron beam evaporating the thin Cr layer. Without breaking vacuum, a 10-nm SiO$_2$ layer is then deposited to protect the Cr from oxidization. Since the SiO$_2$ layer is very thin, it has a marginal effect on the optical properties of the Salisbury screen (see Figure 2). For the two-and three-layer surfaces, we repeat the single-layer steps in succession, but increase the interfacial SiO$_2$ layer thickness to 50 nm, to prevent the solvent from penetrating into underlying polyimide layers during spin coating.

The free-standing Salisbury screen was fabricated from a 1.8 micron sheet of CP1 polyimide, onto which we evaporated 100 nm Cr as the back reflector. For ease of handling, the CP1 sheet was supplied on a polypropylene backing sheet. Following the first evaporation step, the membrane was glued to an acrylic frame, then the backing sheet was removed. On the opposite side, we evaporated 2 nm Cr and 10 nm SiO$_2$.

**Optical measurements.** We obtained the emissivity of our samples by measuring their specular reflectance over wavelengths from 2 to 35 microns with the J. A. Woollam IR-VASE infrared spectroscopic ellipsometry system. Because the metallic back reflector is opaque at all wavelengths, we may calculate absorption (and thus emissivity) directly from the reflectance measurements. We also measured the reflectance with a Nicolet iS50 FTIR coupled to a continuum microscope with a 100 µm spot size, over wavelengths from 2.5 to 15 microns. The FTIR microscope illuminates and collects from the surface of the sample with a Cassegrain lens within an angular range from about 15 to 35 degrees, therefore the obtained emissivity is averaged over that angular range. Nevertheless, because the emission from these structures is not very sensitive to angle, we observed that the results of the measurements with the FTIR correspond very well in all cases to the results of the reflection measurements with the ellipsometer at 30 degrees.

**Simulations.** Full-wave electromagnetic simulations were performed using in-home codes based on the transfer matrix method and the codes based on the Fourier modal method available online [37].

## 4. Conclusions

Ultralight layered metasurfaces have been designed and fabricated, exhibiting 300 K emissivity up to 0.9 in the surface normal direction (0.85 hemispherical). The total thickness of these structures is only 20–50% of the design free-space wavelength, and their areal mass is less than 10 g/m$^2$. We attribute the high spectral emissivity of these metaphotonic structures to different phenomena including phononic resonances in its dielectric layers and transfer of energy of quantum-confined hot electrons in the metallic particulates to the vibrons in the adjacent polyimide.

We also fabricated a free-standing Salisbury screen with an emissivity of 0.6 and weighing only 3 g/m$^2$. These metasurfaces have substantial mechanical flexibility, low outgassing [38], resistance to ultraviolet radiation and atomic oxygen, and the extremely low areal mass density of only a few g/m$^2$, making them of considerable interest for space-based and other ultralight flexible technology applications.

## Acknowledgement

We acknowledge financial support from the Northrop Grumman Corporation; E.C.W and H.A.A. were partially supported by the DOE "Light-Material Interactions in Energy Conversion' Energy Frontier Research Center under grant DE-SC0001293. A.N. acknowledges support from the Swiss Science National Foundation. We acknowledge Tom Tiwald of J. A. Woollam Co. for analyzing the ellipsometry measurements of the polyimide layers, and Lynn Rodman of Nexolve for providing materials and guidance in fabricating the thin polyimide layers. We also thank Mark Kruer, George Rossman, Laura Kim, Victoria Chernow, Michelle Sherrot and Will Whitney for assisting with emissivity measurements; Dagny Fleischman, Rebecca Glaudell, Cristofer Flowers and Rebecca Saive for their support during the fabrication and measurements; and Colton Bukowsky and Krishnan Thyagarajan for technical discussions.

Supplementary information

1. The polarized spectral emissivity of the Salisbury screen

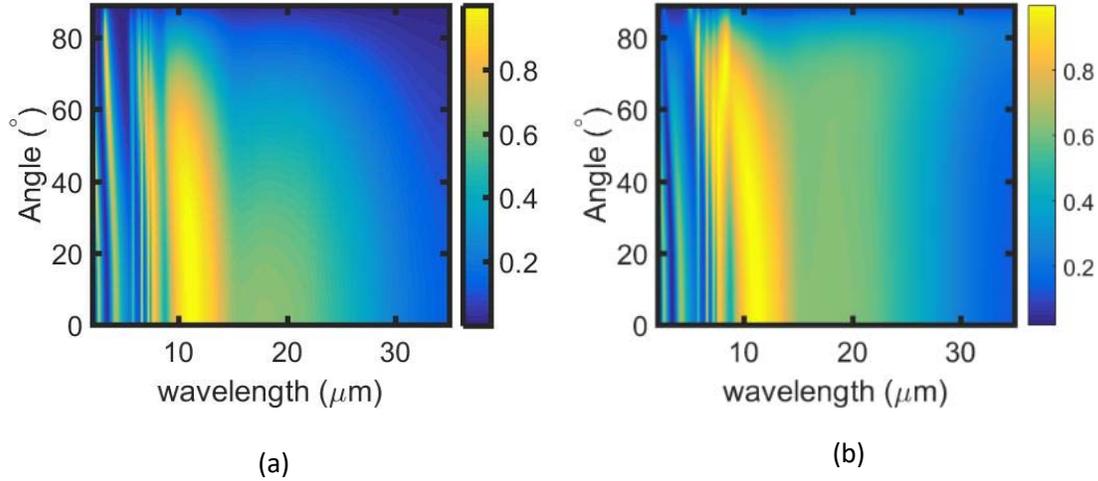

(a)    (b)

*Figure S1- Spectral emissivity of the Salisbury screen versus the emission angle and the wavelength for (a): TE-polarized emission, and (b): TM-polarized emission.*

Figure S1 shows the angle dependence of the spectral emissivity in both polarizations for the Salisbury screen. The width of the pronounced Fabry-Perot resonance is a few microns due to the subwavelength thickness of the dielectric spacer layer. As the incidence angle increases, the resonance experiences a blue shift in both polarizations, but these changes are not very significant due to the large width of the resonance. In TM polarization, the discontinuity of the electric field at the Cr/CP1 interface leads to a Brewster-like effect, which appears in Figure S1 (b) as higher emissivity over a broad angular range for long wavelengths up to almost 25 microns. Also the phononic resonance of polyimide near 8 microns gets amplified at large angles. These two phenomena lead to a considerable amount of absorption in the polyimide layer and maximize the TM-polarized absorption at around 70 degrees. Therefore, the TE and TM polarized results combined give rise to nearly constant absorption and emissivity over a very wide angular range.

## 2. Measured emissivities

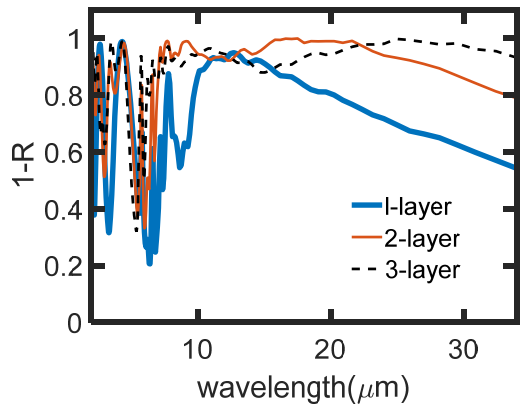
(a)

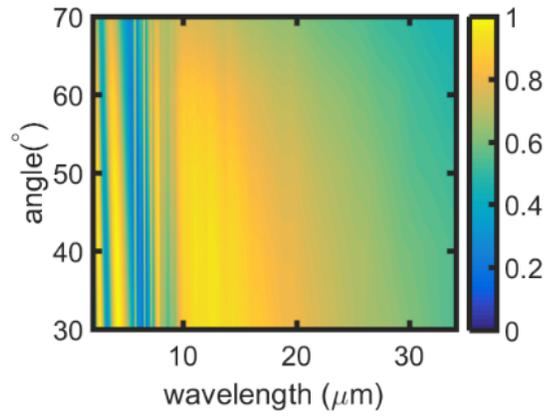
(b)

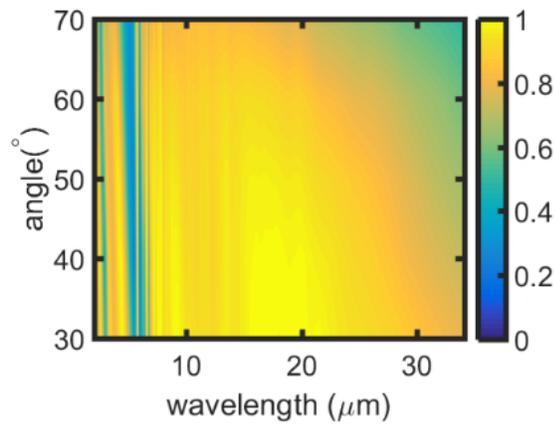
(c)

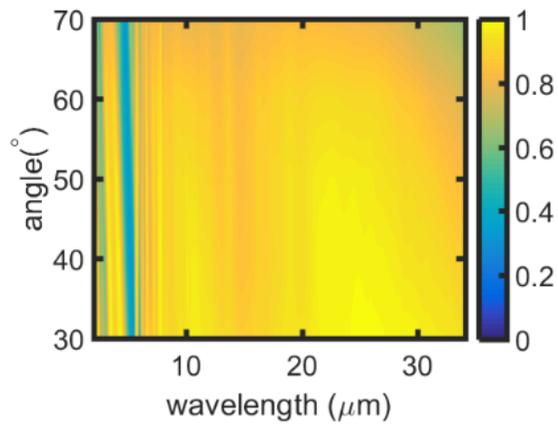
(d)

*Figure S2- Infrared spectral emissivity of the fabricated structures, inferred from reflection measurements. (a) at 30 degrees, (b): 1-layer structure, (c): 2-layer structure, (d): 3-layer structure.*

The free-standing CP1 foil

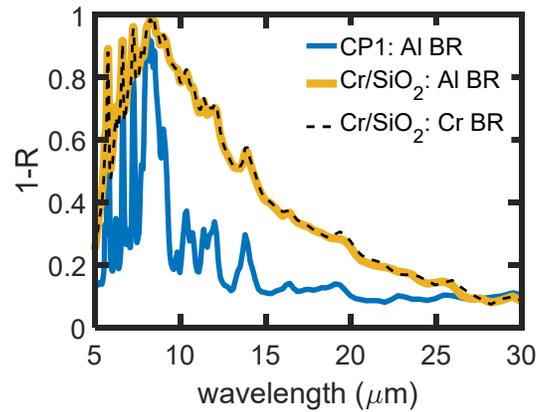

*Figure S3- The spectral emissivity (inferred from specular reflectance measurements at 30 degrees) of CP1 film with (a): Al back reflector and nothing on top, (b): Al back reflector and 2 nm Cr and 10 nm SiO₂ on top, and (c) Cr back reflector and 2 nm Cr and 10 nm SiO₂ on top. The emissivity of the free-standing foils is notably less than that of those fabricated on Si wafers because the thickness of the CP1 layer is not ideal for the free-standing foils.*